\documentclass[12pt,preprint]{aastex}
\usepackage{psfig}

\newcommand{\HI}{{\ion{H}{1}}}

\newcommand{\kms}{$\,$km$\,$s$^{-1}$}

\newcommand{\kmsMp}{km s$^{-1}$ Mpc$^{-1}$}

\def\HI{H{\,\small I}}

\def\emph#1{{\sl #1}}
\newcommand{\ltsima} {$\; \buildrel < \over \sim \;$}
\newcommand{\gtsima} {$\; \buildrel > \over \sim \;$}
\newcommand{\lta} {\lower.5ex\hbox{\ltsima}}
\newcommand{\gta} {\lower.5ex\hbox{\gtsima}}

\slugcomment{\date}
\shorttitle{Fast \HI\ outflow in 3C~293}
\shortauthors{Morganti et al.}
\received{2003 June 17}
\begin{document}

\title{Fast  outflow  of neutral hydrogen \\
in the radio galaxy 3C~293
\thanks{Based on observations  with the Westerbork Synthesis Radio Telescope
(WSRT)}}

\author{R.  Morganti, T.A.  Oosterloo}
\affil{Netherlands Foundation for Research in Astronomy, Postbus 2,
NL-7990 AA, Dwingeloo, The Netherlands}
\email{morganti@astron.nl, oosterloo@astron.nl}

\author{B.H.C. Emonts, J.M. van der Hulst}
\affil{Kapteyn Astronomical Institute, RuG, Landleven 12, 9747 AD,
Groningen, NL}
\email{emonts@astro.rug.nl, j.m.van.der.hulst@astro.rug.nl}
\and
\author{C.N.  Tadhunter}
\affil{Dep. Physics and Astronomy,
University of Sheffield, Sheffield, S7 3RH, UK}
\email{C.Tadhunter@sheffield.ac.uk}

\date{Received ...; accepted ...}

\begin{abstract}

We report the detection of very broad \HI\ absorption against the central
regions of the radio galaxy 3C~293. The absorption profile, obtained with the
Westerbork Synthesis Radio Telescope, has a full width at zero intensity of
about 1400 \kms\ and most of this broad absorption ($\sim 1000$ \kms) is
blueshifted relative to the systemic velocity.  This absorption represents a
fast outflow of {\sl neutral} gas from the central regions of this
AGN. Possible causes for such an outflow are discussed. We favour the idea
that the interaction between the radio jet and the rich ISM produces this
outflow. Some of the implications of this scenario are considered.

\end{abstract}
\keywords{galaxies: active - galaxies: individual: 3C~293 - ISM: jets and  
outflow - radio lines: galaxies}

The physical and kinematical conditions of the gas surrounding an
active galactic nucleus (AGN) offer key diagnostics for understanding
the impact of the nuclear activity on the inter-stellar medium
(ISM). As the gas is likely to be found in different phases, the study
of all possible phases (atomic, molecular and ionized) is crucial to
obtain a complete view of the processes occurring, in particular in
the inner few kpc around the nucleus.

The picture of these regions provided by the ionized gas is often
quite complex. For example, fast gas outflows are now detected (from
optical, UV and X-ray observations) in a wide range of AGNs, from
Seyfert galaxies to quasars (see e.g.\ Crenshaw et al.\ 2000, 2001;
Aoki et al.\ 1996; Turnshek 1986 and refs. therein).  Unambiguous
evidence for outflows have been recently found also in radio galaxies
from studies of the ionized gas (Tadhunter et al.\ 2001, Holt et
al.\ 2003). These outflows can be produced by different and highly
energetic phenomena, such as interaction of the radio plasma with the
ISM as well as nuclear and/or starburst winds.  Apart from telling us
about the way the energy is released by the active nucleus, the
feedback mechanisms originating by these outflows are considered to be
critical in regulating the growth of the central black-hole (BH) and
possibly explaining, e.g., the correlation between BH and galaxy bulge
properties (see e.g.\ Silk \& Rees 1998).

Neutral hydrogen is another important diagnostic to investigate the
physical conditions and the kinematics of the gas in the central
regions of AGNs. So far, observations of \HI\ absorption in radio
galaxies (but also in Seyfert galaxies) have been often interpreted as
the neutral gas being mainly associated with circum-nuclear tori or
kpc-scale disks (see e.g.\ the case of Cygnus~A, Hydra~A or NGC~4261,
Conway \& Blanco 1995, Taylor 1996, van Langevelde et al.\ 2000).  In
these cases, the neutral gas is supposed to trace relaxed/settled
structures and its kinematics is found to be less extreme than that
of the ionized gas. However, recent observations indicate that such an
interpretation cannot always be applied and that also for the neutral
gas the situation can be much more complex (Morganti 2002 and
refs therein) with, for example, interaction between the radio jet
and the neutral gas (see e.g.\ the case of the Seyfert galaxies IC~5063
and Mrk~1; Oosterloo et al.\ 2000, Omar et al.\ 2002 respectively).

Here we present the discovery of an even more extreme case. New \HI\
data of the radio galaxy 3C~293, obtained with the broad band system
now available at the Westerbork Synthesis Radio Telescope (WSRT),
reveal the presence of a fast outflow of {\sl neutral} gas in the
central regions of this radio source.  This is the first case found in
a radio galaxy. This result shows that, despite the extremely
energetic phenomena occurring near an AGN - including the powerful
radio jet as in the case of 3C~293 - some of the outflowing gas
remains, or becomes again, neutral. This result is giving new and
important insights on the physical conditions of the gaseous medium
around an AGN.

\section{The broad \HI\ absorption from the WSRT observations}

The WSRT observations of 3C~293 were performed at four epochs (August
2002, December 2002, January 2003 and March 2003) using the new
backend that allows a broad observing band (20~MHz) with a large
number of channels (1024) so that a wide velocity range is covered
($\sim 4000$ \kms) while high velocity resolution ($\sim 4$
\kms) can be maintained.  We used different central frequencies in the
different observations (1359.2 MHz in the August and December
observations, 1361.4 and 1357.9 MHz for the remaining two) to
eliminate systematic errors due to bandpass uncertainties.

Each observation was about one hour in duration.  Over this time
interval we verified that the bandpass stability of the system (using
a calibrator before and after the observation of 3C~293) is better
than 1 in $10^4$ in each observation.  The data were calibrated and
reduced using the MIRIAD package. The final r.m.s.\ noise is 0.86 mJy
beam$^{-1}$ with a velocity resolution of 9 \kms\ (after Hanning
smoothing). The peak of the continuum emission is $\sim 3.8$ Jy. The
final spectrum is thermal-noise limited with a spectral dynamic range
of about $2\times 10^{-4}$.

The \HI\ spectrum, obtained from the combination of all data, is shown
in Fig.\,1. In addition to the strong and relatively narrow \HI\
absorption already known (Haschick \& Baan 1985, Beswick et al.\
2002), we detect a new very broad component.  This broad component is
also detected in each of the four observations when analysed
separately.

The broad \HI\ absorption (clearly seen in the zoom-in of Fig.\,1),
has a full-width at zero intensity (FWZI) of $\sim 1400$ \kms and, if
fitted with a gaussian,  FWHM$\sim 850$ \kms.  This absorption extends
from $\sim -1000$ \kms\ to +400 \kms\ (i.e.\ from $\sim 12450$ \kms\ to
$\sim 13850$ \kms) compared to the systemic velocity - taken from the
literature as 13500 \kms.

The structure of the continuum emission of 3C~293 at 21 cm at the
resolution of the WSRT is shown in Fig.\ 1 (from Emonts et al.\ in
prep.). The typical size of the WSRT synthesised beam is about 10
arcsec, corresponding at $\sim 9.2$ kpc at the distance of 3C~293 (for
H$_{\circ }$ = 71 \kmsMp).  Despite the short duration of our observations,
we can verify that all the detected \HI\ absorption (including the new
broad component) is coming from the central region, because the
fan-beam is perpendicular to the radio source structure.  It is worth
mentioning that the central region, unresolved with the WSRT beam, is
in fact relatively complex when observed at higher spatial
resolution. A two-sided jet structure and a flat-spectrum core were
observed by Akujor et al.\ (1996). The brighter structure detected is
an hot-spot-like region situated around 1.5 arcsec (1.4 kpc) from the
core.

The large velocities associated with the broad \HI\ absorption are very
unlikely to be associated with gravitational motion and instead indicate a
fast gas outflow in the nuclear regions.  This is clearly different from the
deep \HI\ absorption that was has been previously studied in detail by, e.g.,
Haschick \& Baan (1985) and Beswick et al.\ (2002). Two gas systems are
believed to produce this relatively narrow part of the \HI\ absorption. One
system, at larger radii, likely represents the neutral counter-part of the
large-scale ionized disk (van Breugel et al.\ 1984) and of the molecular gas
disk (Evans et al.\ 1999). The second system is a more inner ring of gas (with
a radius of at least 600 pc) rotating around the active nucleus (Beswick et
al.\ 2002).

Finally, the broad \HI\ absorption component is very shallow, with a typical
optical depth of only $\sim 0.15$ \%, assuming that it uniformly covers the
central radio source (i.e.\ the covering factor is 1), The column density of
the \HI\ is $\sim 2 \times 10^{20}\ T_{\rm spin}/100$ K cm$^{-2}$. This is
likely to be a lower limit to the true column density as the $T_{\rm spin}$
associated with such a fast outflow can be as large as few 1000~K (instead of
100~K which is more typical of the cold, quiescent \HI\ in galaxy disks) and
the covering factor may well be smaller than one. Thus, the true column
density will likely be in excess of $10^{21}$ cm$^{-2}$.

\section{The origin of the broad \HI\ absorption}

The discovery of the broad, blueshifted \HI\ absorption indicates that
fast outflows of {\sl neutral} gas can exist near AGNs. The obvious
question is, therefore, which of the mechanisms capable of producing
gas outflows can also account for gas that remains, or becomes again,
neutral.  All the different mechanisms thought to produce fast
outflows (interaction between the radio plasma and the ISM, nuclear
and/or starburst winds) can be at work in the nuclear regions of a
radio galaxy like 3C~293. We shall therefore consider each of them
separately.

3C~293 is known to have a particularly rich ISM. A large amount of
molecular gas has been detected ($\sim 1.5 \times 10^{10} M_{\sun}$,
Evans et al.\ 1999) and this galaxy is also a bright far-IR source.
The high concentration of molecular gas in the central
few kpc and the distorted optical morphology of the galaxy has led to
suggestions that 3C 293 has been involved in a recent gas-rich galaxy-galaxy
interaction (Heckman et al. 1986; Evans et al. 1999).
A relatively young stellar population component is also observed in
optical spectra (Robinson 2001). The presence of a starburst wind is,
therefore, quite conceivable.  The age of the young stellar population
component has been estimated between 0.4 and 2.5 Gyr. Thus, the
neutral outflow would have to be a ``fossil'' starburst-driven wind
from the strong starburst that may have occurred of the order of 1 Gyr
ago. Given this condition, it is not clear whether the
starburst-driven outflow would survive to the present day and whether
it would be still seen against the central regions. In fact, following
Heckman, Armus \& Miley (1990), the size of the gas shell produced by
a starburst wind of this age will be very large and not limited
anymore to the nuclear region. We, therefore, consider the effect of
such wind unlikely to produce the fast \HI\ outflow that we
see against the nuclear regions of 3C~293.

More promising mechanisms to explain the outflow are those related to
the effects of the AGN activity.  As mentioned in the introduction,
fast gas outflows have been detected in a number of Seyfert galaxies.
In some of these objects (in particular where no correlation between
the radio structures and the peculiar velocities of the ionized gas
has been found) acceleration of the gas due to radiation and/or wind
pressure has been proposed. Dopita et al.\ (2002) have considered in
detail the case of dusty narrow-line regions that are radiation
pressure dominated. This mechanism may be able to explain the overall
kinematics of the gas (including the fast outflow) in the case of
NGC~1068. In this scenario, if the column density is high enough, the
gas clouds are ionisation bounded and some neutral gas can be present
in the outflow.

However, unlike NGC~1068, the gas in the centre of 3C~293 appears to
have very low ionisation and has very faint emission lines (in
particular [\ion{O}{3}]5007\AA, Gelderman \& Whittle 1994). This could
indicate that the AGN in 3C~293 is relatively weak. Based on the
observation of the emission lines, this is, however, quite difficult
to quantify given the strong obscuration by the dusty environment
around the centre. It is nevertheless clear that the ionisation of
this galaxy appears to be low. 

We can alternatively use the far-IR luminosity of 3C~293.  Even
assuming that the far-IR emission is all due to re-radiated quasar
light - rather than re-radiated starburst light - the FIR luminosity
(from IRAS observations, Golombek et al.\ 1988) is $\log {L}_{60\mu
m}/{L_{\odot}} = 10.08$.  This is at the lower end of the luminosity
for quasars (Neugebauer et al.\ 1986), therefore indicating, as the
low ionisation level, that the nucleus of 3C~293 does not have a
particularly powerful AGN in the centre.  It seems therefore unlikely
that there is enough energy in the UV radiation field from the nucleus
to be able to accelerate the gas to such large velocities as observed
in the \HI\ profile.

The last, and perhaps more likely, possibility is that the outflow is
the driven by the interaction between the radio plasma and the
ISM. Although the WSRT observations cannot resolve the complex nuclear
structure of 3C~293 (see above) and therefore exactly locate where the broad
\HI\ is occurring, we know that some interaction between the radio
plasma and the ISM is taking place.  Evidence was found that the
broadest optical lines are seen in coincidence with the region of the
most intense radio emission, which is a radio hot-spot about 1.4 kpc
east of the core, as determined by VLBI observations (Akujor et al.\
1996).  New optical spectra (see Fig.\ 2, from Emonts et al.\ in
prep.)  show that in this location the optical emission lines contain
a broad component that is very similar to the broad \HI\ absorption.
This suggests that also the \HI\ absorption is  coming from this
region. However, this will need to be confirmed with higher resolution
\HI\ observations.

Also for this scenario, the central question is how {\sl neutral} gas can be
associated to the fast outflow.  A possible model is that the radio plasma jet
hits a (molecular) cloud in the ISM.  As a consequence of this interaction,
part of the gas is ionized and its kinematics is disturbed by the shock.  Once
the shock has passed, part of the gas may have the chance to recombine and
become neutral, while it is moving at high velocities.  To understand whether
this scenario could be feasible, it is worth considering the model proposed
for the evolution of clouds in radio galaxy cocoons as they are overtaken by a
strong shock wave (Mellema, Kurk \& R\"ottgering 2002).  This model predicts
that, as the shock runs over a cloud, a compression phase starts because the
cloud gets embedded in an overpressured cocoon. The shock waves start
travelling {\sl into} the cloud and the cloud fragments with the fragments
moving at high velocities.  They find that the cooling times for the dense
fragments are very short  (few times $10^2$ years) compared to the lifetime of
the radio source and that the excess of energy is quickly radiated away. This
results in the {\sl formation of dense, cool and fragmented structures at high
velocities}.  It would be interesting to explore the parameter space of such
models in more detail in order to see whether, for the conditions in 3C~293,
an outflow of neutral gas with the high velocities observed can be produced.

As indirect support to this hypothesis, it is worth mentioning that in
the only other case of broad blueshifted \HI\ absorption (of 700 \kms\
FWZI) studied in detail so far (the Seyfert galaxy IC~5063,
Oosterloo et al.\ 2000), the \HI\ absorption is coincident with the
brighter radio lobe where also the most kinematically disturbed
ionized gas is observed. This supports the idea of jet/cloud
interaction as most likely mechanism in this Seyfert galaxy.  A
detailed high-resolution study to investigate whether or not a spatial
coincidence exists between the broad component seen in ionized and
neutral gas and the features seen in radio continuum is needed to
further clarify whether IC~5063 and 3C~293 are indeed similar.

\section{Final remarks} 

Broad, blueshifted \HI\ absorption, as reported here for 3C~293, is only very
rarely seen.  As mentioned above, the only other object studied in some detail
where broad blueshifted absorption has been found is the Seyfert galaxy
IC~5063 (Oosterloo et al.\ 2000).  This could well be due to a technical
bias. In order to detect the broad and shallow \HI\ absorption, the radio
source has to be quite strong while also broad-band (i.e.\ 16 MHz or more)
observations with very high spectral dynamic range are necessary, conditions
that are not satisfied by most observations available in the literature.
Indeed, most Seyfert galaxies are too weak in the radio for detecting broad,
shallow \HI\ absorption.

The detection of such a component is more likely, given their stronger radio
emission, in radio galaxies and a systematic search is now in progress.  We
have indeed recently found two other candidate radio galaxies (4C~12.50 and
3C~305) where similarly broad and blueshifted \HI\ absorption may be
occurring.

It is intriguing that 4C~12.50 and 3C~305, as well as 3C~293, are classified
as ``starburst'' radio galaxies, i.e.\ show evidence of a
relatively young stellar population component. Such galaxies appear
to be among the best candidates for detecting \HI\ in general (see e.g.\
Morganti et al.\ 2001).  This might be due to the richer ISM that
characterises radio galaxies in this stage of their evolution, with the rich
ISM possibly resulting from a recent merger. It is then perhaps not surprising
that fast gas outflows are also produced in this rich ISM by the young radio
source. More observations are needed to study how common neutral gas outflows
are and whether they are associated with particular kinds of AGN.

Following these results, a tantalising connection can be made with the
high-$z$ radio galaxies. Outflow phenomena have been detected in many
high-redshift radio galaxies. In many cases, asymmetric Ly$\alpha$
profiles suggest the presence of blueshifted absorbing gas (likely
neutral hydrogen, Dey 1999, van Ojik et al.\ 1997, de Breuck et al.\
1999). Additionally, complex gas kinematics is also observed in a
large fraction of high-$z$ radio galaxies (van Ojik et al.\ 1997).
There is clear evidence for the presence of large amounts of
cold gas and, in general, for the presence of a rich gaseous
environment in radio galaxies at high redshift (see e.g.\ van Breugel
2000 for a review).  Strong interactions between the radio plasma and
the medium are therefore expected to be very important. Because of
this, similar processes as observed in 3C~293 are likely to be even
more common in these high-$z$ systems. Understanding the physics of
fast gas outflows and the conditions for which part of the outflowing
gas is neutral, is thus also quite  relevant for understanding
high-$z$ radio galaxies.

\begin{acknowledgements}

The WSRT is operated by the ASTRON (Netherlands Foundation for
Research in Astronomy) with support from the Netherlands Foundation
for Scientific Research (NWO).
\end{acknowledgements}

{} 
\clearpage

\begin{figure*}
\centerline{\psfig{figure=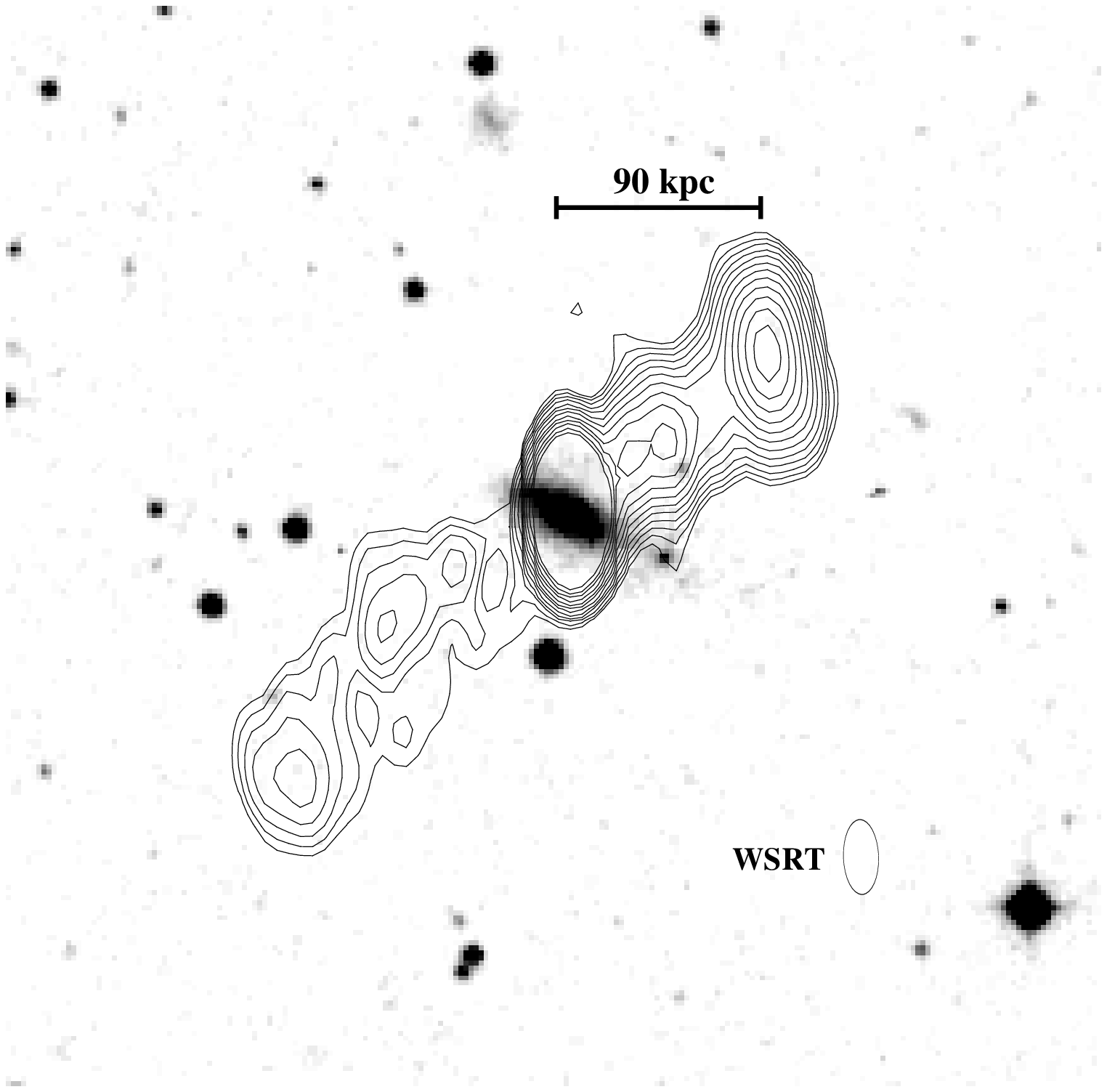,angle=0,width=6cm}\psfig{figure=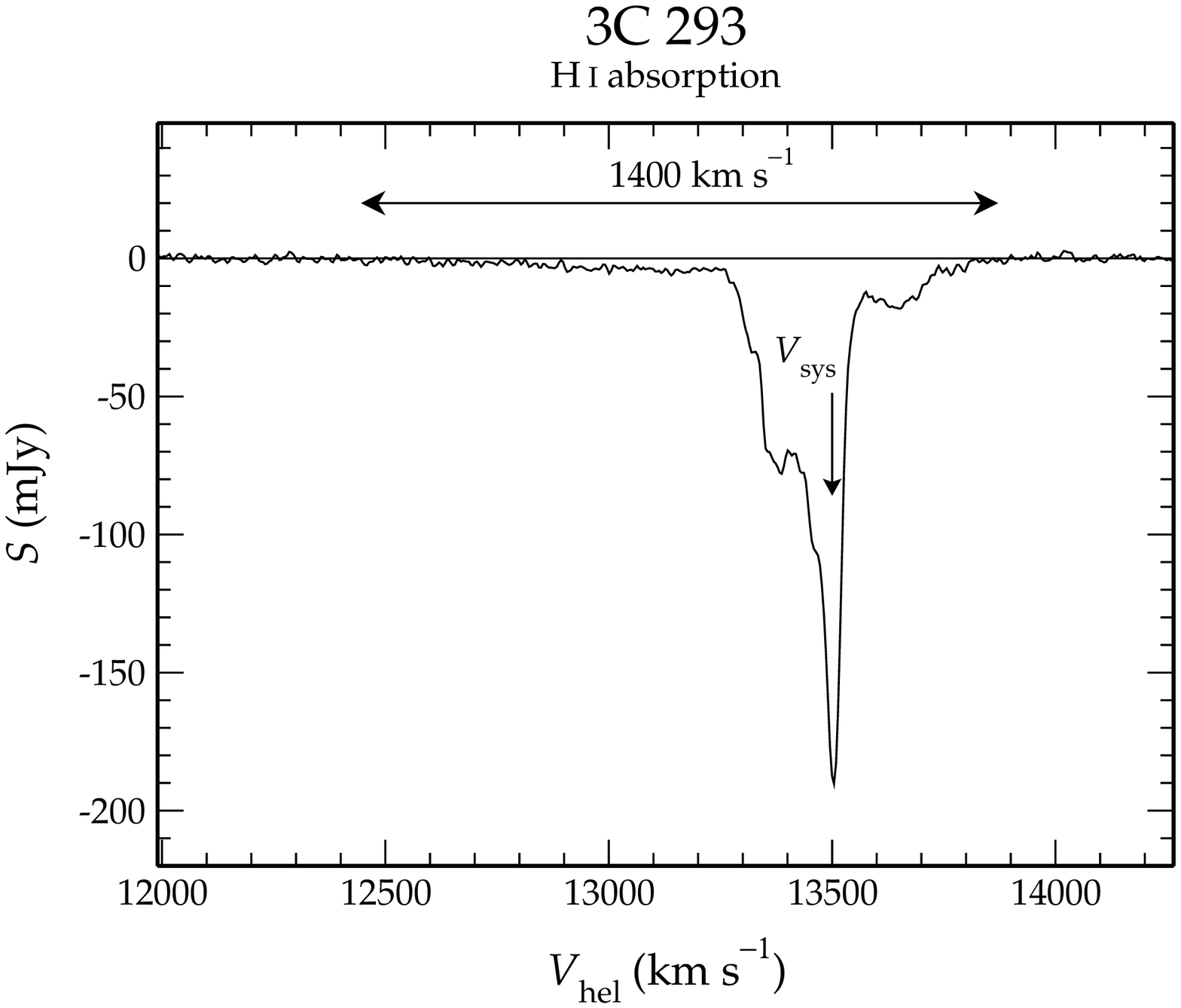,angle=0,width=6cm}\psfig{figure=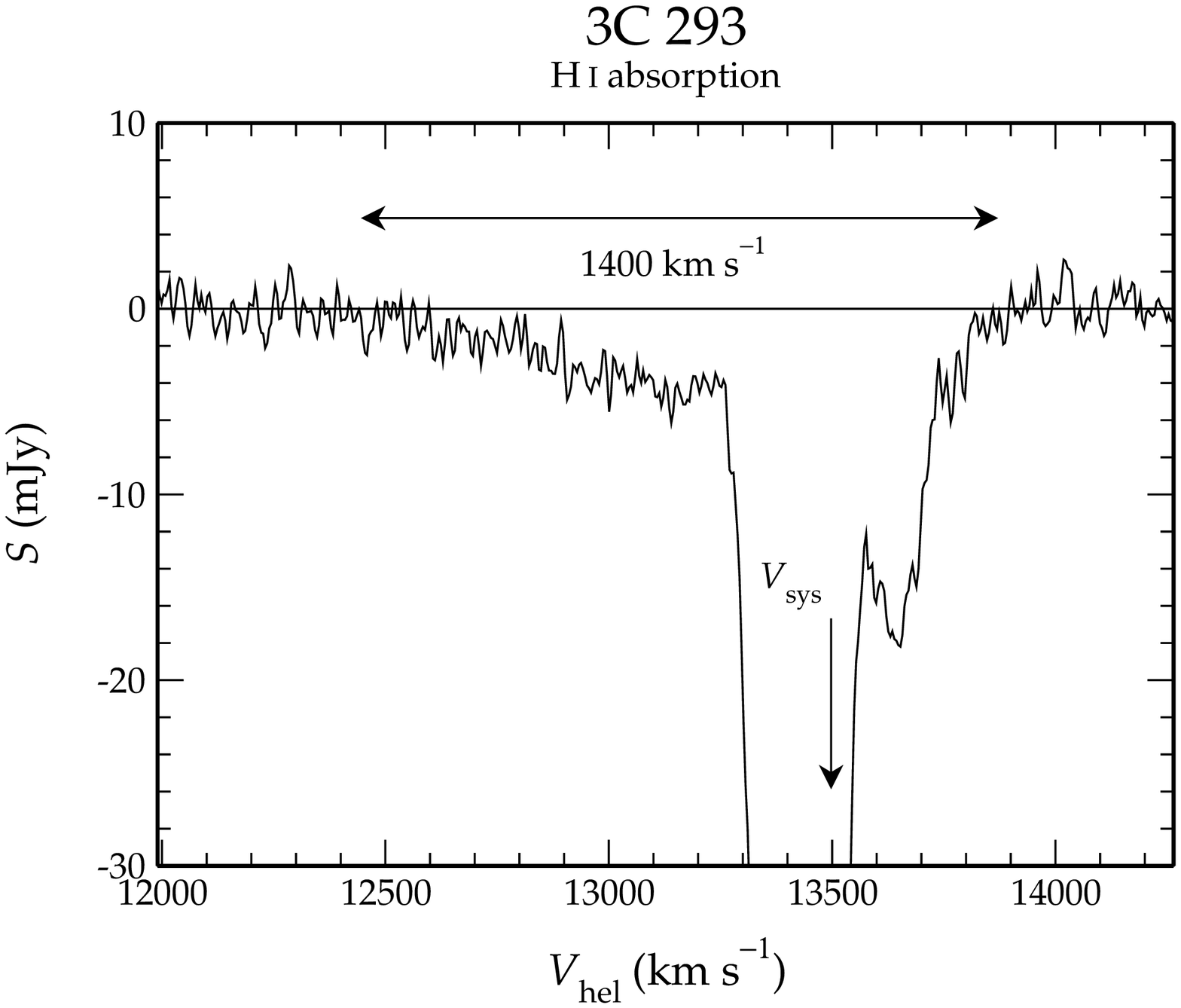,angle=0,width=6cm}}
\caption{{\sl (Left)} Continuum image of 3C~293 at the resolution of WSRT 21-cm
observations (Emonts et al. in prep). {\sl (Middle)} The \HI\ absorption spectra
with a zoom-in ({\sl Right}) to better show the new detected broad \HI\
absorption. The spectra are plotted in flux (mJy) against optical heliocentric
velocity in \kms.} 
\label{fig:prof_a}
\end{figure*}

\clearpage

\begin{figure*}
\centerline{\psfig{figure=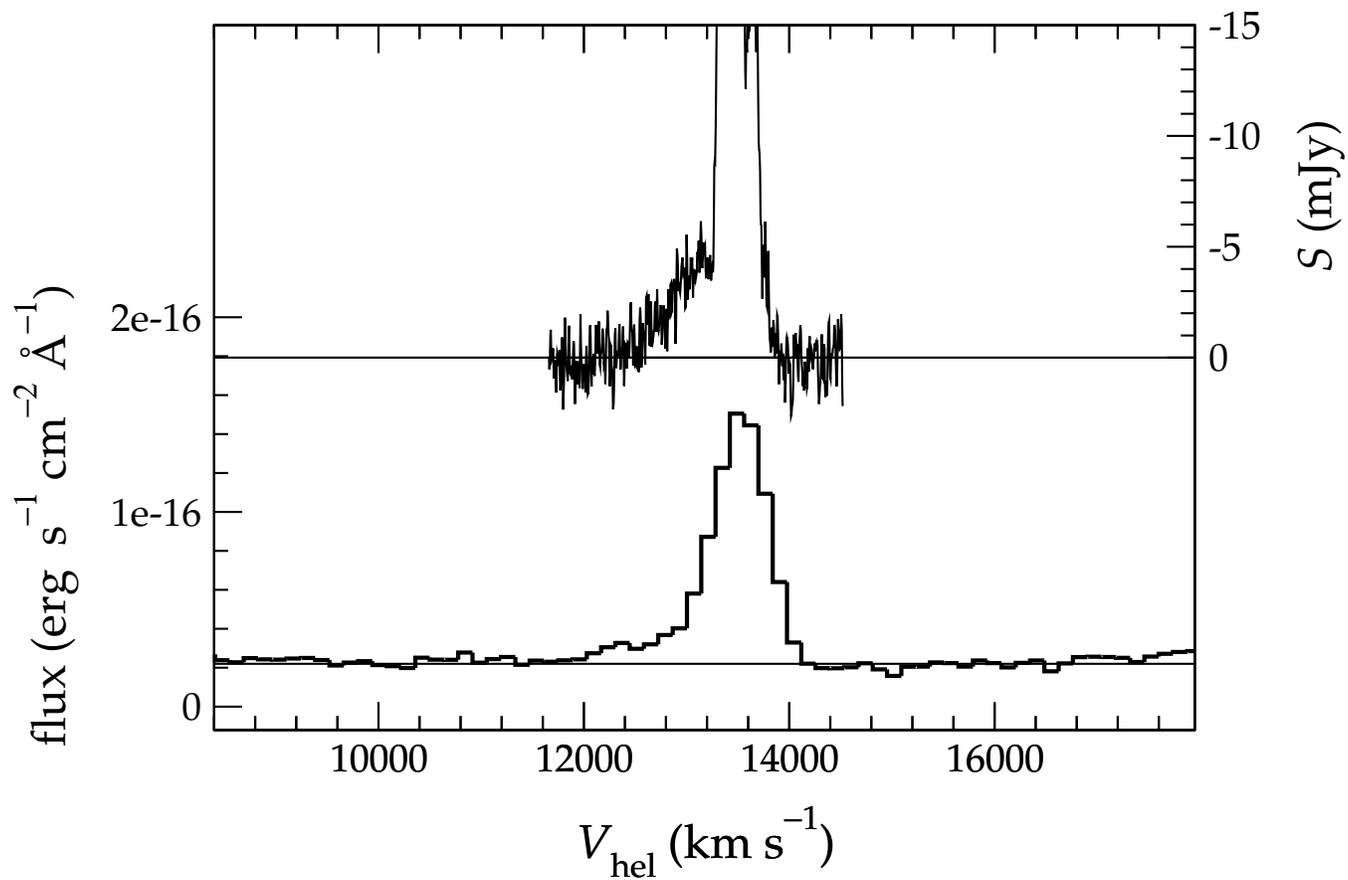,angle=0,width=18cm}}
\caption{Comparison between the  \HI\ absorption (top) and the [OII]3727\AA\ 
(bottom, from Emonts et al. in prep.) profiles.  The
similarity of the broad, blueshifted wing in the two profiles is
evident.}
\label{fig:prof_b}
\end{figure*}

\end{document}